\documentclass{bioinfo}
\copyrightyear{2014}
\pubyear{2014}

\usepackage{fixltx2e}
\usepackage{natbib}
\begin{document}
\firstpage{1}

\title[Artifacts in variant calling]{Towards Better Understanding of Artifacts in Variant Calling from High-Coverage Samples}
\author[Li]{Heng Li}

\address{Broad Institute of Harvard and MIT, 7 Cambridge Center, Cambridge, MA 02142, USA}

\history{Received on XXXXX; revised on XXXXX; accepted on XXXXX}
\editor{Associate Editor: XXXXXXX}
\maketitle

\begin{abstract}

\section{Motivation:} Whole-genome high-coverage sequencing has been widely
used for personal and cancer genomics as well as in various research areas.
However, in the lack of an unbiased whole-genome truth set, the global error
rate of variant calls and the leading causal artifacts still remain unclear
even given the great efforts in the evaluation of variant calling methods.

\section{Results:} We made ten SNP and INDEL call sets with two read mappers
and five variant callers, both on a haploid human genome and a diploid genome
at a similar coverage. By investigating false heterozygous calls in the haploid
genome, we identified the erroneous realignment in low-complexity regions and
the incomplete reference genome with respect to the sample as the two major
sources of errors, which press for continued improvements in these two areas.
We estimated that the error rate of raw genotype calls is as high as 1 in
10--15kb, but the error rate of post-filtered calls is reduced to 1 in
100--200kb without significant compromise on the sensitivity.

\section{Availability:} BWA-MEM alignment and raw variant calls are available at http://bit.ly/1g8XqRt
Scripts and miscellaneous data at https://github.com/lh3/varcmp

\section{Contact:} hengli@broadinstitute.org

\end{abstract}

\section{Introduction}

Since the sequencing of the first personal genome~\citep{Levy:2007uq} and in
particular the first genomes sequenced with the Illumina
technologies~\citep{Bentley:2008cr,Wang:2008nx}, resequencing has been widely
used for personal and cancer genomics~\citep{Watson:2013aa}, for the discovery
of de novo mutations associated with Mendelian diseases~\citep{Bamshad:2011aa},
for the reconstruction of human population history~\citep{Li:2011ij} and for
the understanding of mutation processes~\citep{Veltman:2012aa,Campbell:2013aa}.
In most of these studies, mapping based SNP/INDEL calling plays a central role.
The accuracy of the calls has a fundamental impact on the biological
interpretation. In this context, various research groups have attempted to
evaluate the performance of variant calling.

The simplest approach to the evaluation of variant calling is to simulate
variants and reads from a reference genome~\citep{Li:2008zr}. However, we are
unable to simulate various artifacts such as the non-random distribution of
variants, dependent errors, incomplete reference genome and copy number
variations. An improved version is to incorporate real variants instead of
using simulated variants~\citep{Talwalkar:2013aa}, but it does not address the
artifacts caused by large-scale effects, either. A better simulation is to take
the reads sequenced from one sample with a finished genome, map them to another
finished genome, call variants and then compare the calls to the differences found by genome-to-genome
alignment~\citep{Li:2008zr}. However, this approach is limited to small haploid
genomes. There are attempts to apply a similar idea to mammalian
genomes~\citep{Li:2013ab,Bolosky:2014aa}, but as the mammalian reference
genomes are frequently incomplete and the whole-genome alignment is imperfect,
such a simulation is still different from realistic scenarios.

The difficulties in simulation have motivated us to focus more on real data. One
simple approach is to thoroughly sequence a small target region with mature
technologies, such as the Sanger sequencing technology, and take the resultant sequence as
the ground truth~\citep{Harismendy:2009aa}. It does not capture large-scale
artifacts, though. Another more commonly used method is to measure accuracy
either by comparing variant calls from different pipelines, or by comparing
calls to variants ascertained with array genotyping or in another
study~\citep{Clark:2011aa,Li:2012fk,Lam:2012aa,Lam:2012fk,Boland:2013aa,Liu:2013aa,Goode:2013aa,ORawe:2013aa,Zook:2014aa,Cheng:2014aa}.
However, array genotyping is biased to easier portions of the genome and may have a higher error rate per assayed site than
the variant calling error rate~\citep{Bentley:2008cr}; simply comparing call
sets would only give us an estimate of the relative accuracy -- if two
pipelines are affected by the same artifact that a third pipeline does
not have, the third pipeline will appear worse even though it is in fact
better.  In addition, comparative studies usually measure the accuracy with
summary statistics such as the fraction of calls present in dbSNP or the
transition-to-transversion ratio.  They do not tell us the wrong sites.

Many studies also experimentally validated typically up to a few hundred
variants with MiSeq or Sanger sequencing or Sequenom genotyping. Nonetheless,
such experiments are biased towards easier regions and may also be subjected
to other artifacts such as on-primer variants and non-specific amplification
(the 1000 Genomes Project analysis subgroup, personal communication).
Calling heterozygotes from Sanger sequence data is also challenging by itself.


In the author's view, it is better to evaluate variant calling by
comparing samples from a pedigree~\citep{Zook:2014aa}, or from the same
individual~\citep{Nickles:2012aa}, including cancer
samples~\citep{Lower:2012aa}. Because we expect to see only tens to hundreds of
somatic mutations or Mendelian errors per genome~\citep{Conrad:2011kx}, most other inconsistencies
are very likely to be errors. However, this method is insensitive to systematic
errors.  If at a locus, a caller finds erroneous heterozygotes in all samples,
the errors wouldn't be identified.

After these efforts, we are still not clear about a basic question: the error
rate of SNP and INDEL calling. Although a few papers give an estimate of one
error per 100--200kb, it is either estimated on easy
sites~\citep{Bentley:2008cr} or not sufficiently backed with published
data~\citep{Nickles:2012aa}. In addition, only a few
works~\citep{Larson:2012aa,Roberts:2013aa,Kim:2013aa} have attempted to
identify the sources of errors. Analyzing systematic errors is even rarer as
most existing evaluation methods hide them.

In this article, we use an exceptional data set, sequencing data from a
haploid human cell line, to evaluate the accuracy of variant calling. As the
vast majority of heterozygous calls are supposed to be errors, we almost know
the ground truth unbiasedly across the whole genome. We are able to pinpoint
errors, investigate their characteristics, experiment filters and get a
reasonable estimate of the error rate, not limited to non-systematic errors. In
addition to the unique data set, our study also differs from many previous ones
in the use of multiple read mappers, unpublished but well developed variant
callers, and caller-oblivious genotyping and filtering.

\begin{methods}
\section{Data Sets and Data Analysis}

\subsection{Data sets}

In this study, we focused on deep Illumina sequencing data from two cell lines,
the CHM1hTERT cell line~\citep{Jacobs:1980aa} and the NA12878 cell line. A
crucial and unusual feature of CHM1hTERT, or briefly CHM1, is that this cell
line is haploid, which suggests that any heterozygous variant calls are errors.
A calling method producing fewer heterozygotes is in theory better. Meanwhile,
to avoid overrating a variant calling method with low sensitivity on
heterozygotes, we also used NA12878 as a positive control.

The entire CHM1 data set (AC:SRP017546) gives over 100-fold coverage. We are
only using 6 SRA runs with the accessions ranged from SRR642636 to SRR642641.
The 6 runs are from the same library, yielding about 65-fold coverage before
the removal of potential PCR duplicates.

We acquired the NA12878 data set (AC:ERR194147) from the Illumina Platinum
Genomes project. The library was constructed without PCR amplification. We are
only using paired-end data, which yields about 55-fold coverage.

\subsection{Alignment and post-alignment processing}

We mapped the CHM1 reads with Bowtie2~\citep{Langmead:2012fk} and
BWA-MEM~\citep{Li:2013aa} and mapped the NA12878 reads with
BWA~\citep{Li:2009uq} in addition to Bowtie2 and BWA-MEM. The detailed command
lines can be found in Table~1. Except in Section~\ref{sec:ref}, we mapped the
reads to hs37d5, the reference genome used by the 1000 Genomes Project in the
final phase.

After the initial alignment, we run Picard's MarkDuplicates on both data sets.
Picard identified 20\% of CHM1 reads as PCR duplicates. For NA12878,
Picard reported 1.5\% of them as PCR duplicates, which are false positives as
the library was constructed without amplification. We did not apply
MarkDuplicates for NA12878 in the subsequent analysis.

For the NA12878 BWA alignment, we also tried GATK's~\citep{Depristo:2011vn}
base quality recalibration (BQSR) and INDEL realignment around INDEL calls from
the 1000 Genomes Project~\citep{1000g:2012aa}. For both
SAMtools~\citep{Li:2011ab} and GATK, the number of calls only differ by 0.1\%,
much smaller than the difference caused by other procedures. We thus did not
apply these steps to other alignments due to the additional computational cost.
It should be noted that although BQSR and INDEL realignment have little effect
on these two high-coverage data sets, it may make difference on low-coverage
data or when the base quality is not well calibrated.

\subsection{Calling SNPs and short INDELs}

We called SNPs and short INDELs with FreeBayes~\citep{Garrison:2012aa}, GATK
UnifiedGenotyper, Platypus, SAMtools and GATK HaplotypeCaller. The command
lines can be found in Table~1. Additional details are as follows.

\begin{table*}
\footnotesize
\processtable{Evaluated mappers and variant callers}
{\begin{tabular*}{\textwidth}{@{\extracolsep{\fill}}lp{1.8cm}lp{12.5cm}}
\toprule
Symbol & Algorithm & Version & Command line \\
\midrule
bt2 & bowtie2 & 2.1.0 & bowtie2 -x \emph{ref.fa} -1 \emph{read1.fq} -2 \emph{read2.fq} -X 500 \\
bwa & bwa-backtrack & 0.7.6 & bwa aln -f \emph{read1.sai} \emph{ref.fa} \emph{read1.fq}; bwa sampe \emph{ref.fa} \emph{read1.sai} \emph{read2.sai} \emph{read1.fq} \emph{read2.fq} \\
mem & bwa-mem & 0.7.6 & bwa mem \emph{ref.fa} \emph{read1.fq} \emph{read2.fq} \\
fb & freebayes & 0.9.9 & freebayes -f \emph{ref.fa} \emph{aln.bam} \\
st & samtools & 0.1.19 & samtools mpileup -Euf \emph{ref.fa} \emph{aln.bam} {\tt \char124} bcftools view -v - \\
ug & UnifiedGenotyper & 2.7-4 & java -jar GenomeAnalysisTK.jar -T UnifiedGenotyper -R \emph{ref.fa} -I \emph{aln.bam} \mbox{-stand\_call\_conf 30} \mbox{-stand\_emit\_conf 10} -glm BOTH \\
hc & HaplotypeCaller & 2.7-4 & java -jar GenomeAnalysisTK.jar -T HaplotypeCaller --genotyping\_mode DISCOVERY -R \emph{ref.fa} -I \emph{aln.bam} \mbox{-stand\_call\_conf 30} \mbox{-stand\_emit\_conf 10} \\
pt & Platypus & 0.5.2 & Platypus.py callVariants --filterDuplicates=1 --bamFiles=\emph{aln.bam} --refFile=\emph{ref.fa} \\
\botrule
\end{tabular*}}{}
\end{table*}

\subsubsection{Resolving overlapping variants} Platypus and SAMtools may
produce many overlapping variants. To avoid overcounting variants, for two
overlapping variants, we always keep the one with the higher variant quality.
We repeated this procedure until no overlapping variants remain.

\subsubsection{Re-calling genotypes} Given the same genotype likelihood,
different callers may produce different genotypes. For example, SAMtools
estimates genotypes assuming the prior of seeing a heterozygote being $10^{-3}$,
but GATK does not apply a prior. GATK is more likely to call a heterozygote
than SAMtools. Genotype calling for a single sample is relatively simple.
To avoid the subtle difference in this simple step complicating the final
results, we re-call the genotypes from genotype likelihoods provided by the
callers.  We multiplied $10^{-3}$ to the likelihood of heterozygotes and then
called the genotype with the maximum likelihood.

Platypus does not give genotype likelihoods for multi-allelic variants. We
kept the reported genotypes in the VCF.

\subsubsection{Decomposing complex variants} Both FreeBayes and Platypus may
report a variant composed of multiple SNPs and/or INDELs. We decomposed such
variants into individual events such that the results are more comparable.
FreeBayes uses a CIGAR string to describe how a complex variant is aligned to
the reference. We extracted SNPs and INDELs from the CIGAR. Platypus does not
report CIGAR. We assumed the variant allele is always left aligned to the
reference allele when decomposing a complex variant.

\subsection{Variant filtering}\label{sec:flt}

All the callers used in this study come with filtering programs or a recommended
set of filters. However, applying caller-specific filters may complicate
comparison and obscure artifacts. We decided to choose several universal
filters applicable to most callers:

\begin{enumerate}

\item Low-complexity filter (LC): filtering variants overlapping with
low-complexity regions (LCRs) identified with the mdust program
(http://bit.ly/mdust-LC), which is a stand-alone implementation of the DUST
algorithm first used by BLAST. In GRCh37, 2.0\% of A/C/G/T bases on autosomes
are identified to be LCRs.

\item Maximum depth filter (DP): filtering sites covered by excessive number of
reads. It should be noted that different callers may define the depth
differently. For example, Platypus apparently only counts reads with
unambiguous realignment. The read depth reported in the Platypus VCF is
noticeably smaller in comparison to other callers.

\item Allele balance filter (AB): filtering sites where the fraction of
non-reference reads is too low.

\item Double strand filter (DS): filtering variants if either the number of
non-reference reads on the forward strand or on the reverse strand is below a
certain threshold. This filter is not applicable to GATK calls as GATK does not
report these numbers. DS has been identified to be an effective filter on
cancer data~\citep{Roberts:2013aa,Kim:2013aa}.

\item Fisher strand filter (FS): filtering sites where the numbers of
reference/non-reference reads are highly correlated with the strands of the
reads. More precisely, we counted the number of reference reads on the forward
strand and on the reverse strand, and the number of non-reference reads on the
forward and reverse strand. With these four numbers, we constructed a 2-by-2
contingency table and used the P-value from a Fisher's exact test to evaluate
the correlation.

\item Quality filter (QU): filtering sites with the reported variant quality
below a threshold.

\end{enumerate}

Among these filters, LC is a regional filter and is entirely independent of
alignment and variant calling. Although DP is computed from called
variants, its effect is usually not greatly dependent on the mapper and the caller,
either. The remaining filters may be dependent of the error models used by the
callers. For example, SAMtools effectively gives a higher weight on variants
supported on both strand; FreeBayes seems to require a variant to be supported
20\% of reads covering the site. The optimal thresholds for the AB, DS and FS
filters are caller dependent.

\subsection{Measuring accuracy}

The CHM1 and the NA12878 data sets share many properties. They are both
sequenced with 100bp Illumina reads to a similar coverage after the removal of PCR
duplicates. The number of called variants per haplotype is also very close,
usually within 1\% difference according to multiple call sets. Under this
observation, it is reasonable to assume the number of heterozygous errors in
NA12878 is also close to the number of heterozygous calls in CHM1. As a result,
we may take $N_h/N_d$ as an estimate of the false positive rate (FPR) of
heterozygotes, and $N_d-N_h$ as a proxy to sensitivity, where $N_h$ is the
number of heterozygous calls in CHM1 and $N_d$ the number in NA12878. This
might be the first time that we can unbiasedly measure FPR
in a whole genome call set.

\subsection{Manual review}

To understand the major error modes, we have manually reviewed more than 200
heterozygous INDELs called by different callers from CHM1. For these sites,
we displayed the alignment with SAMtools' tview alignment viewer to get a sense
of the alignment quality, obvious positional biases and the complexity of the
reference genome. We often extracted reads in regions around the INDELs,
extending to flanking regions with high complexity to eyes. We assembled the
extracted reads with fermi~\citep{Li:2012fk} version 1.1 and mapped the
assembled contigs back to the reference genome with BWA-MEM. Fermi tries to
preserve heterozygotes. If the INDELs are truly heterozygous, we will typically
see two contigs covering a site, one for each allele. We used the local
assembly as an orthogonal approach to validate heterozygous calls.


\end{methods}

\section{Results}

\begin{figure*}[!ht]
\includegraphics[width=.5\textwidth]{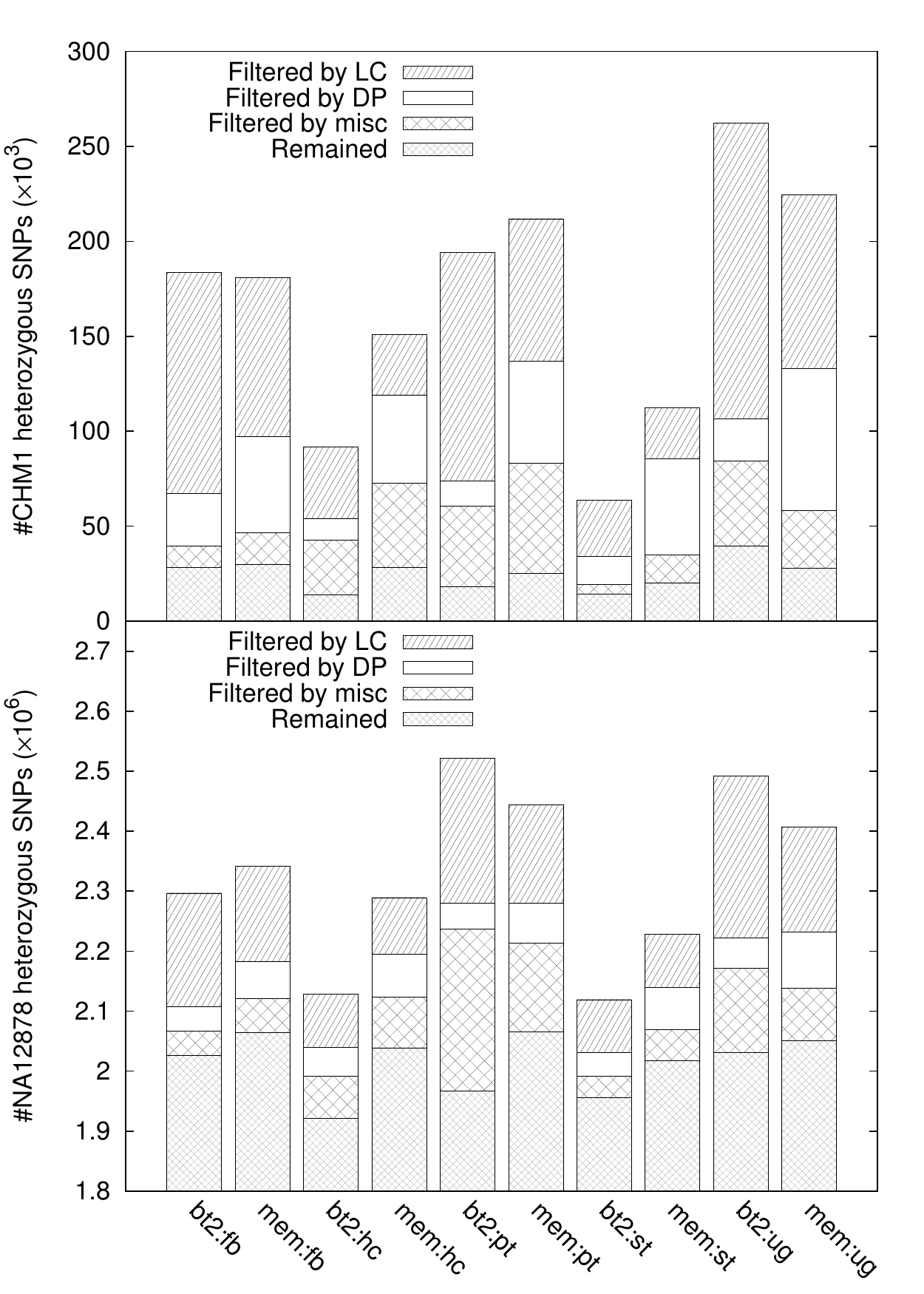}\includegraphics[width=.5\textwidth]{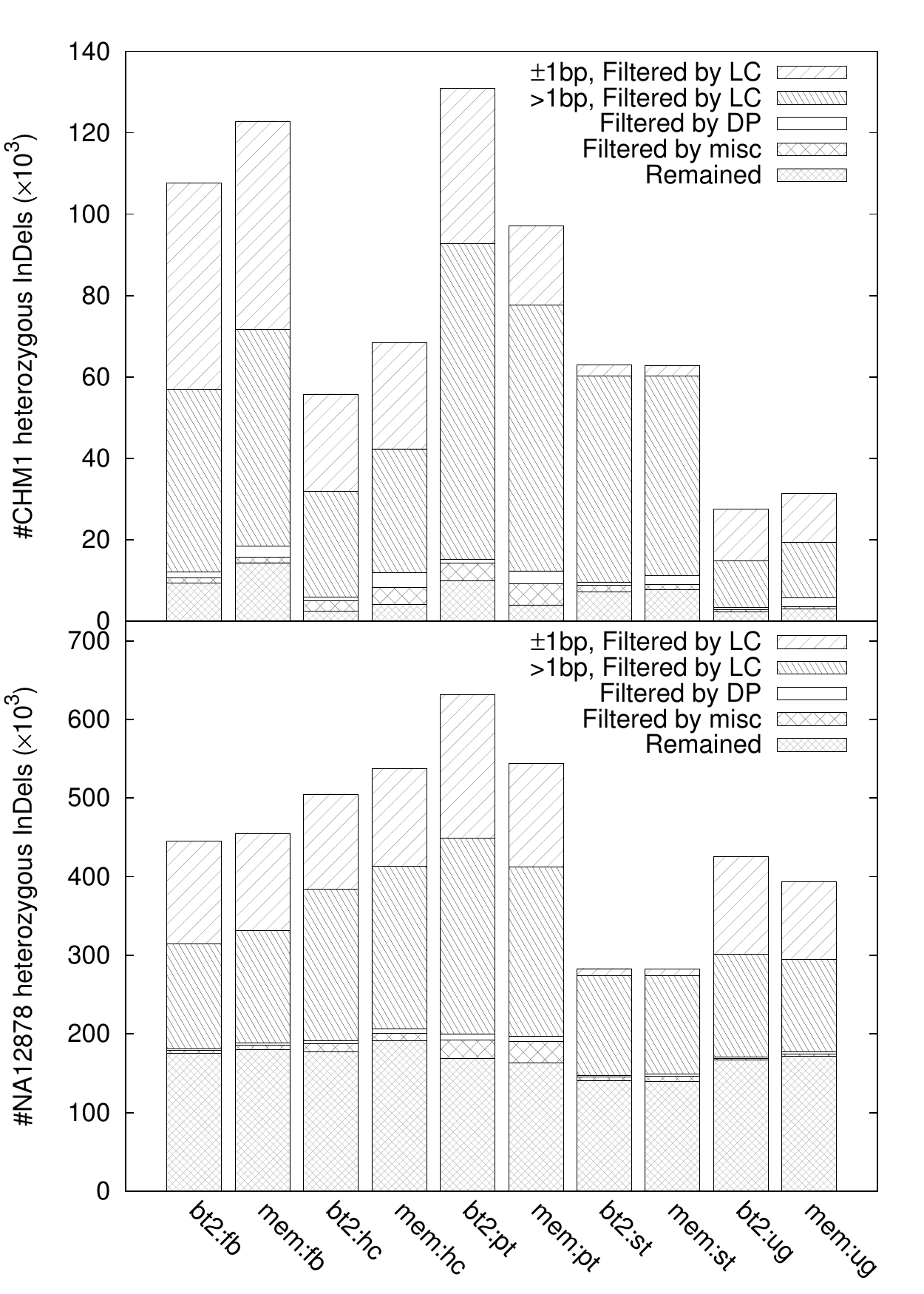}
\caption{Effect of filters. Low-complexity filter (LC): not overlapping
low-complexity regions identified by the DUST algorithm. Maximum-depth filter
(DP): read depth below $d+3\sqrt{d}$, where $d$ is the average read depth.
Miscellaneous filter (misc) includes three filters: allele balance above 30\%,
variants supported by non-reference reads on both strands and Fisher strand
P-value is above 0.01. Filters are applied in the order of LC, DP and misc,
with DP applied to variants passing LC, and misc applied to variants passing
both LC and DP. For each call set, the total height of the bar gives the
number of raw variant calls with the reported quality in VCF no less than 30.
Note that the Y-axes are scaled differently.}\label{fig:hist}
\end{figure*}

When studying the effect of filters on variant calling, we initially applied
the filters independently on each call set. However, when presenting the
results in the following, we applied the filters in an order, with a filter
applied later depending on the filters applied before it. We did this for
clarity and to highlight filters having major effects. Figure~\ref{fig:hist}
overviews the breakdown of various filters across multiple call sets.
If we consider that there might be true heterozygotes in CHM1 potentially due
to somatic mutations, call sets generally have an error rate approximately 1 in
100--200kb (i.e. 15000--30000 false heterozygotes per genome) after filtering.

\subsection{Checking the ploidy of CHM1}

\begin{figure}
\includegraphics[width=.24\textwidth]{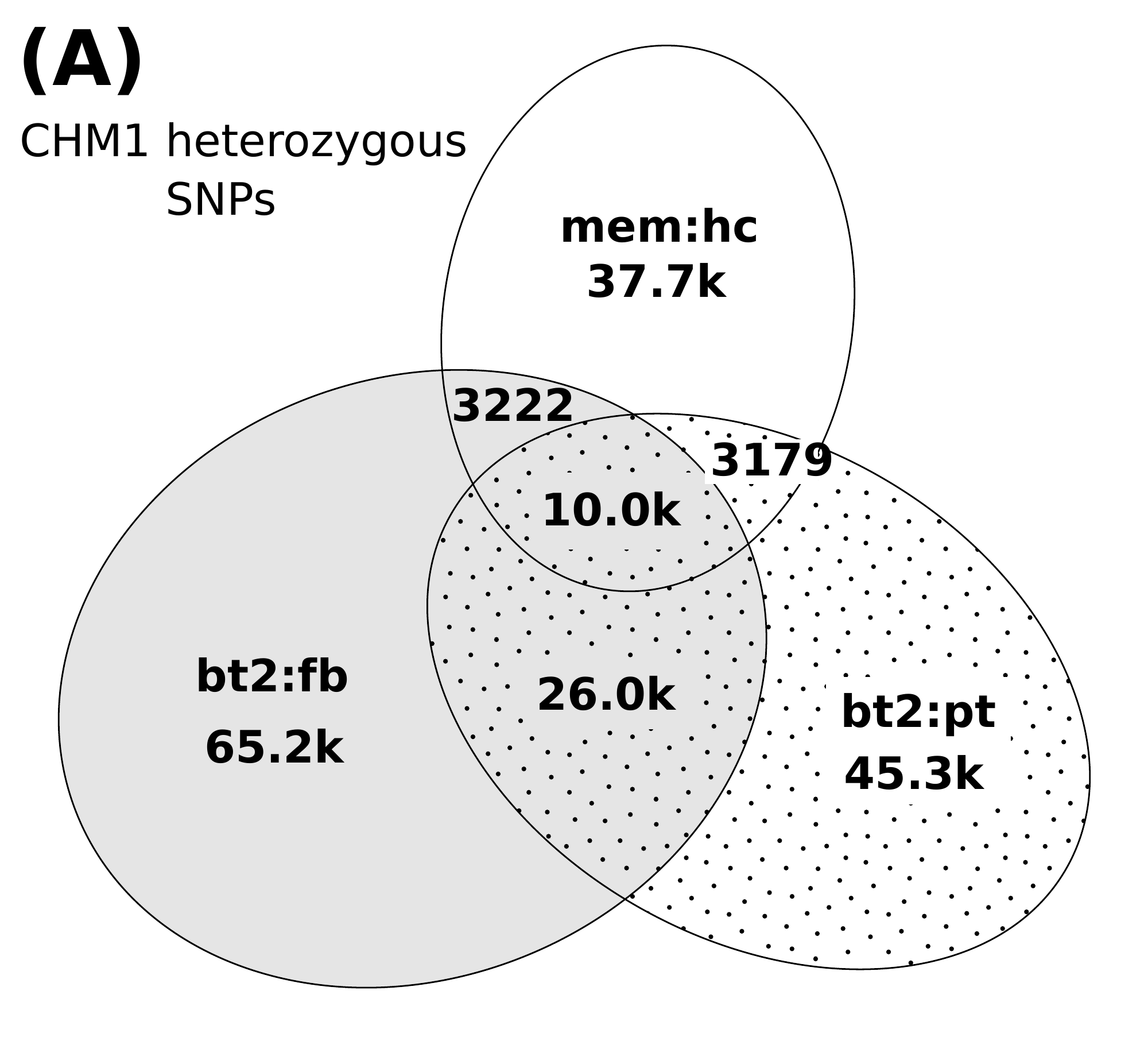}\includegraphics[width=.24\textwidth]{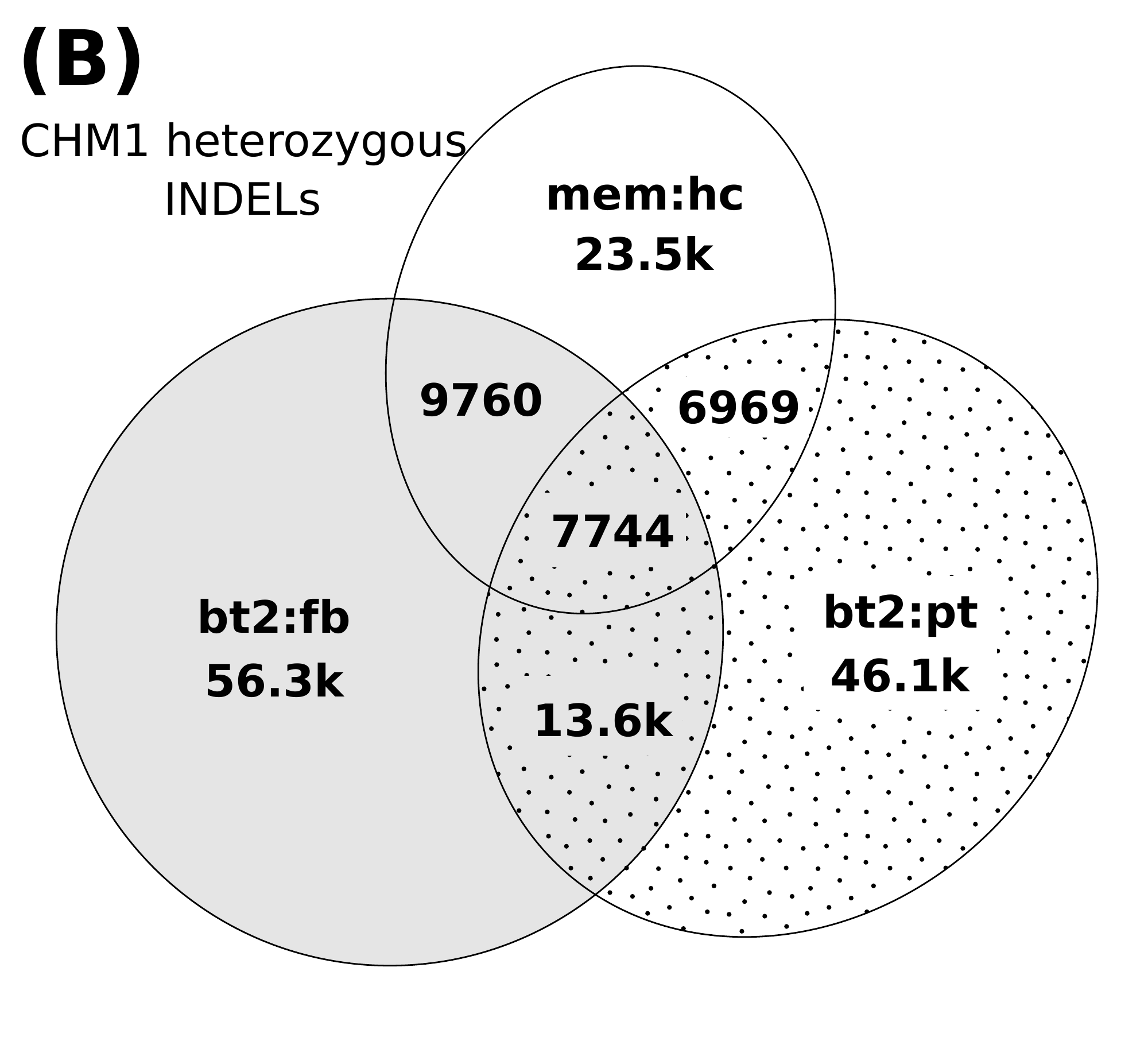}
\caption{Relationship between CHM1 heterozygous call sets. Raw variant calls were filtered
with: variant quality no less than 30, allele balance above 20\%, Fisher strand
P-value above 0.001 and maximum read depth below $d+4\sqrt{d}$, where $d$ is
the average read depth. (A) Relationship between heterozygous SNP call sets. Two SNPs are
considered the same if they are at the same position. (B) Relationship between heterozygous
INDEL call sets. Two filtered INDELs are said to be \emph{linked} if the 3'-end
of an INDEL is within 20bp from the 5'-end of the other INDEL, or vice versa.
An INDEL \emph{cluster} is a connected component (not a clique) of linked
INDELs. It is possible that in a cluster two INDELs are distant from each other
but both overlap a third INDEL.  Venn's diagram shows the number of INDEL
clusters falling in each category based on the sources of INDELs in each
cluster. 15\% of SNPs and 91\% of INDELs in the 3-way intersections overlap
low-complexity regions.}\label{fig:venn-CHM1}
\end{figure}

Although the CHM1hTERT cell line is supposed to be haploid, we may still see
heterozygous variant calls potentially because: a) the cell line is not truly
haploid; b) there are somatic mutations in the cell line; c) there are
library construction and sequencing errors~\citep{Robasky:2014aa}, which ought
to be considered by the calling algorithms; d) mapping or variant calling
algorithms have flaws. In this study, we are focusing on c) and d), but first
we should make sure heterozygotes resulted from a) and b) occur at a much lower
rate.

We note that if the sample submitted for sequencing is not haploid either due
to biological artifacts or massive somatic mutations, a large number of
heterozygotes should be evident from the sequencing data and get called by all
callers. In contrast, if heterozygotes are mostly caused by sequencing errors
or algorithm artifacts, due to the differences in algorithm and error modeling,
callers will call a subset of errors with different characteristics, which will
result in low consistency between call sets. The small call set intersection in
Figure~\ref{fig:venn-CHM1} suggests the latter is the case.

We also manually reviewed tens of heterozygotes called by multiple callers both
on the data in this study and on Illumina data generated from other libraries
(AC:SRR642626--SRR642635 and AC:SRR642750), which were mapped with the original BWA
algorithm~\citep{Li:2009uq} by the 1000 Genomes Project analysis group.
Reviewing the read evidence using an alignment viewer, it appears that more than half of the SNPs are real.
Most of these SNPs have averaged read depth, non-overlap with known segmental duplications~(http://bit.ly/eelabdb)
and are not associated with known error-prone motifs in Illumina sequencing~\citep{Nakamura:2011aa}.
On the other hand, many INDELs in low-complexity regions look like systematic errors called by
all callers (see also Section~\ref{sec:lc}).  We speculate there may be 5--20k
heterozygotes in CHM1 with strong alignment support from multiple Illumina
libraries. It is hard to get a more accurate estimate or to further tell the
sources of these heterozygotes with the data we are using. As we were writing up
this work, Pacific Biosciences released deep resequencing data for the CHM1
cell line. It could be used to isolate errors caused by the Illumina
sample preparation and sequencing. However, mapping and variant calling from
PacBio human data is still in the early phase. We decided to leave out the comparison
to the PacBio data for now.

Anyway, even if we assume the variant calls in the intersection are all present
in the CHM1hTERT cell line, we should still be able to measure an error rate up
to one error per 170kbp (=3Gbp/17.7k). Given that there are 10 times more raw
heterozygous calls in NA12878 than CHM1 (Figure~\ref{fig:hist}), it seems likely that CHM1
heterozygotes are likely errors from major sequencing/calling artifacts.


As a side technical note, we applied milder filters in
Figure~\ref{fig:venn-CHM1} in comparison to Figure~\ref{fig:hist}. We found the
intersection between call sets often becomes smaller with more stringent
thresholds because stringent thresholds reduce the sensitivity in different
aspects of call sets and amplify the subtle differences between calling algorithms. In
addition, in Figure~\ref{fig:venn-CHM1}B, we were clustering INDELs
within 20bp from each other. Increasing the distance threshold to 100bp only
changed the numbers slightly.

\subsection{The low-complexity filter}\label{sec:lc}

\begin{figure}[!htb]
\includegraphics[width=.24\textwidth]{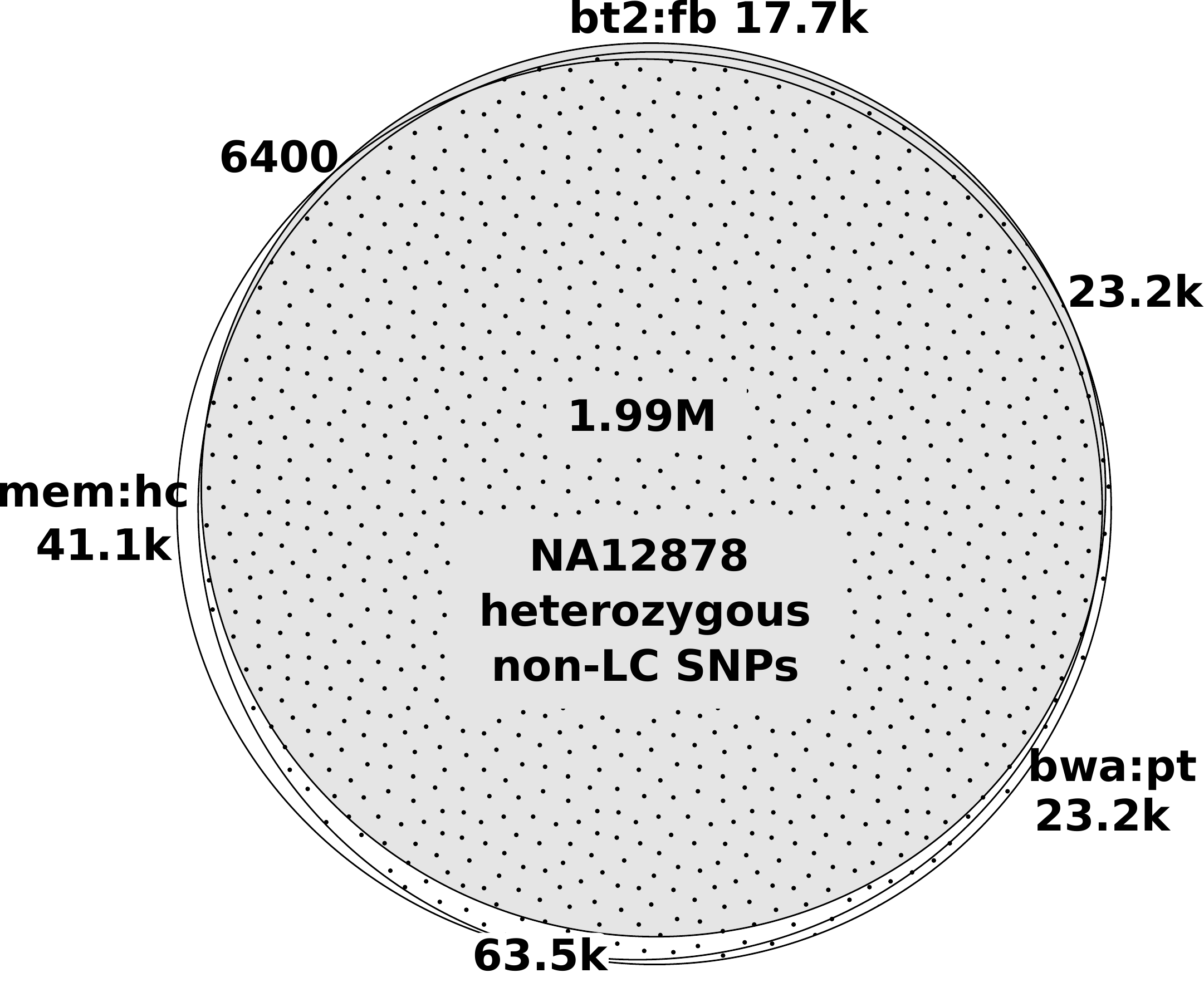}\includegraphics[width=.24\textwidth]{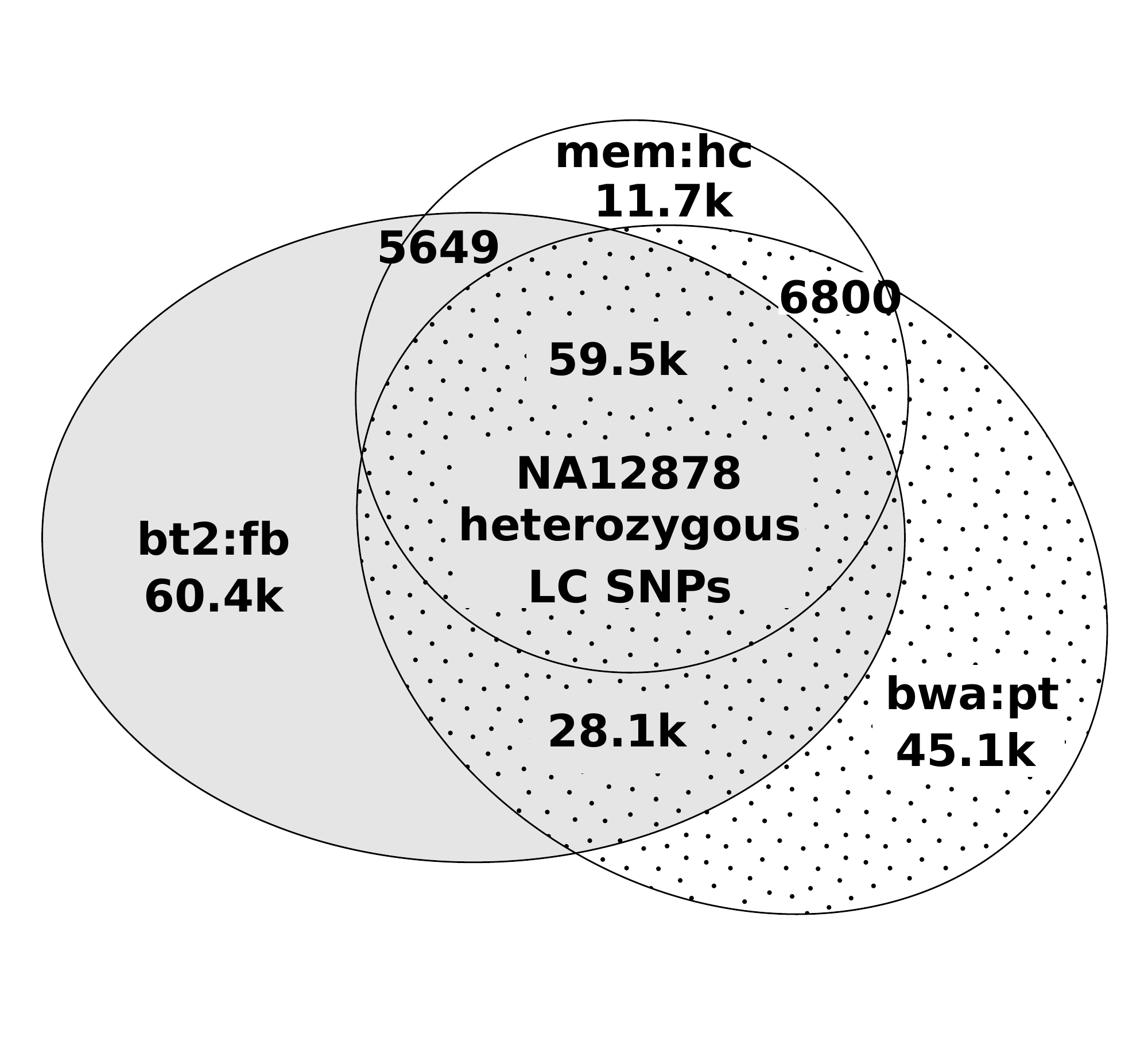}\\
\includegraphics[width=.24\textwidth]{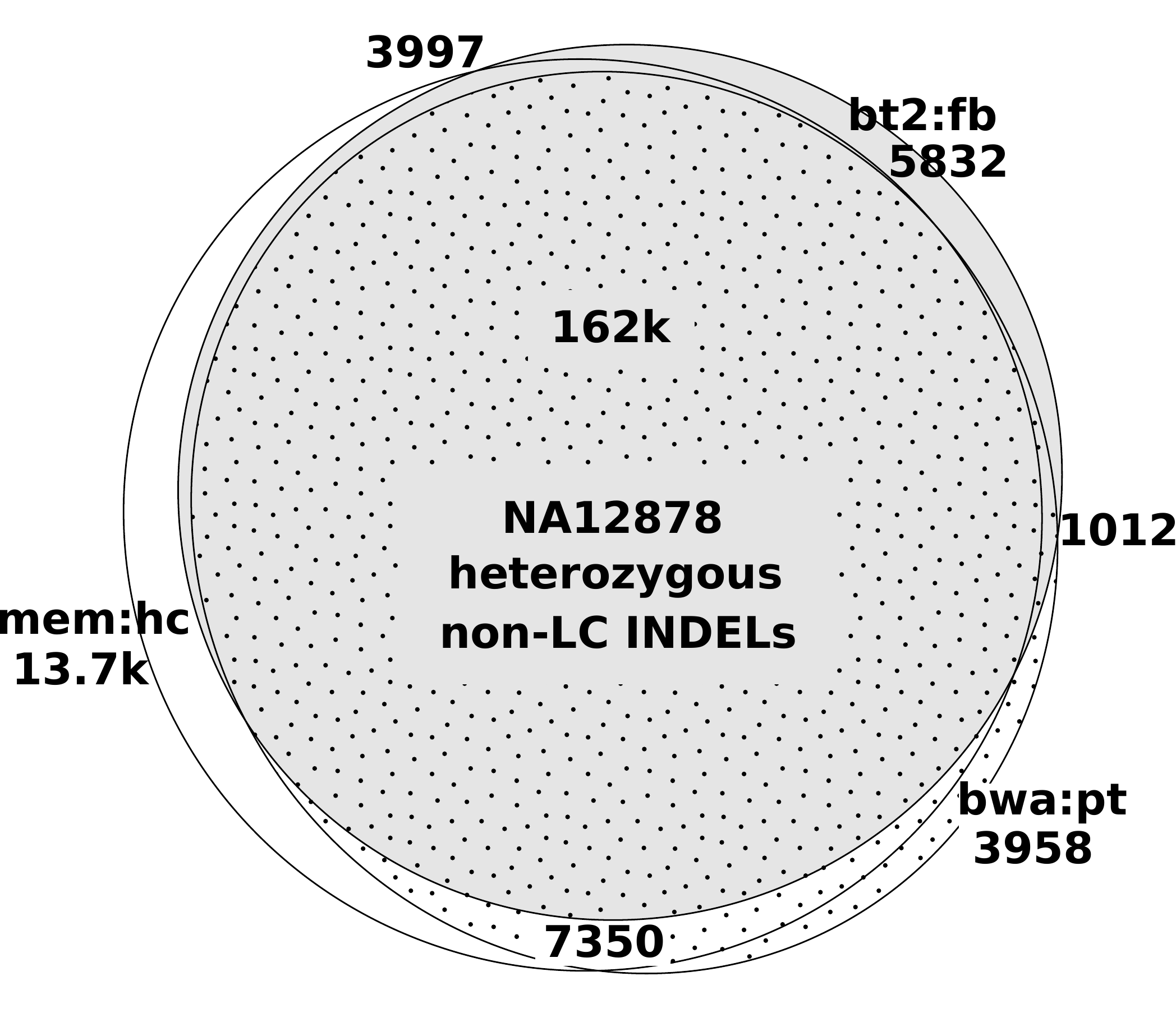}\includegraphics[width=.24\textwidth]{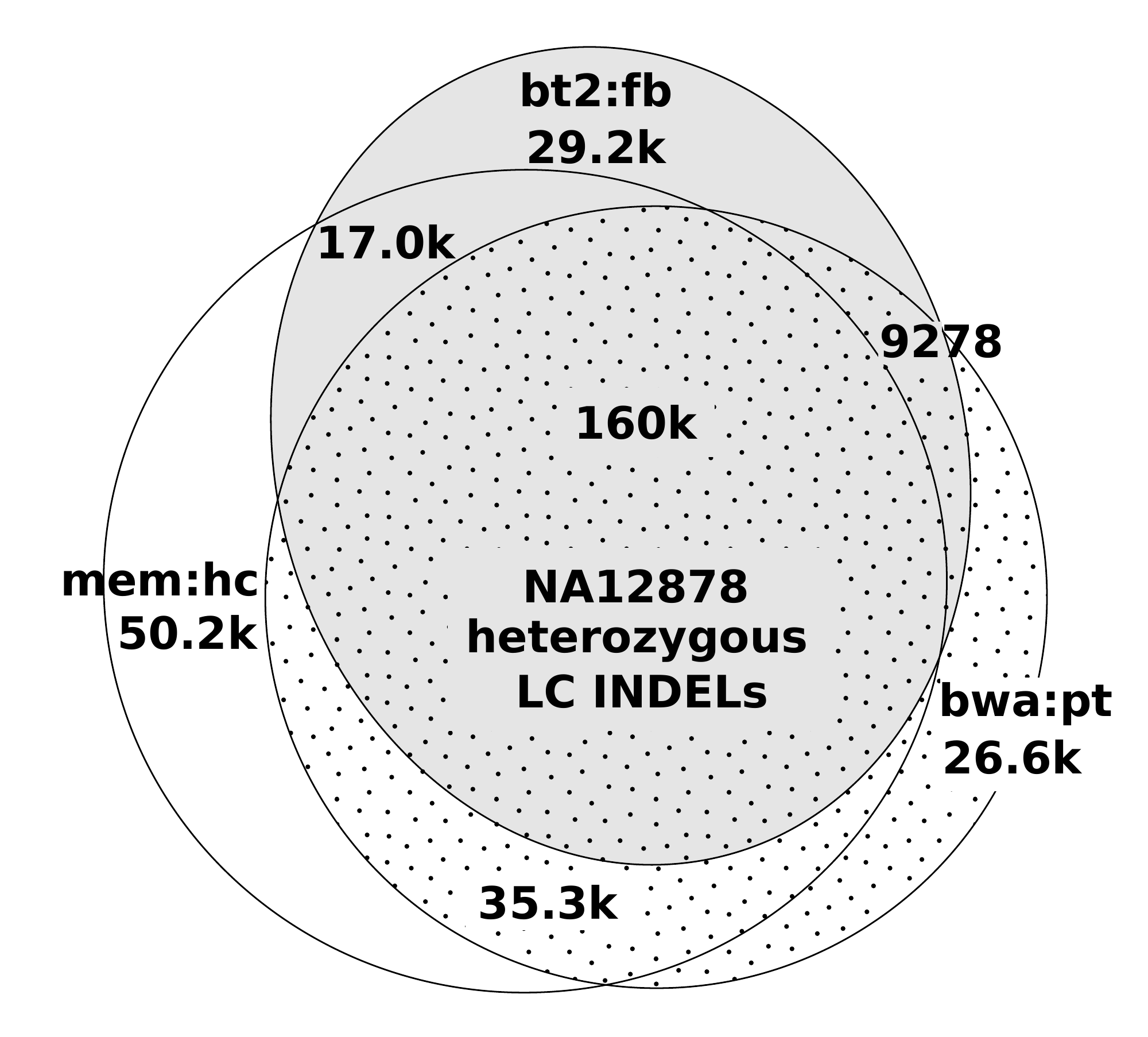}
\caption{Relationship between NA12878 heterozygous call sets.}\label{fig:venn-NA12878}
\end{figure}

On CHM1, low-complexity regions (LCRs), 2\% of the human genome, harbor
80--90\% of heterozygous INDEL calls and up to 60\% of heterozygous SNPs
(Figure~\ref{fig:hist}). Recall that if we let $N_h^{GL}$ be the number of CHM1
heterozygous INDELs in LCRs and $N_d^{GL}$ the number of NA12878 heterozygous
INDELs in LCRs, $N_h^{GL}/N_d^{GL}$ estimates the FPR of heterozygotes. The FPR
in LCRs is ranged from 10\% to as high as 40\% depending on call sets. With a
similar estimator, the FPR of heterozygous INDELs outside LCRs is much lower, about 1--8\% depending on call sets.
We have also tried lobSTR~\citep{Gymrek:2012aa}. It called 65k
heterozygous INDELs from microsatellites, still yielding a high FPR. To
understand why errors are enriched in LCRs, we reviewed over 100 sites and
identified two major sources of INDEL genotyping errors: potential PCR errors
and realignment errors.

\subsubsection{Potential PCR amplification errors}
PCR errors are known to be responsible for many INDEL
errors in long homopolymer runs~\citep{1000g:2012aa}. On CHM1, we have observed many apparent 1bp
heterozygous INDELs (Figure~\ref{fig:hist}) inserted to or deleted from long
poly-A or poly-T runs, which may be due to PCR errors. Although most callers
deploy advanced models for homopolymer INDELs, they are calling vastly
different number of 1bp heterozygous INDELs. It is still not clear to us that
we can model PCR errors well. Maybe the most effective solution
is to avoid PCR in sample preparation.

Potential PCR errors are not the only error source. On the PCR-free NA12878 data
the call set intersection in LCRs is noticeably smaller than in high-complexity
regions (Figure~\ref{fig:venn-NA12878}), which suggests the presence of other
error sources in LCRs. In addition, PCR errors introduced during sample preparation are believed to affect
SNPs to a lesser extend. The small intersections between CHM1 heterozygous SNP
call sets (Figure~\ref{fig:venn-CHM1}), and between PCR-free SNP call sets in
LCRs (Figure~\ref{fig:venn-NA12878}) should be caused by other types of
errors.

\begin{figure*}
\includegraphics[width=\textwidth]{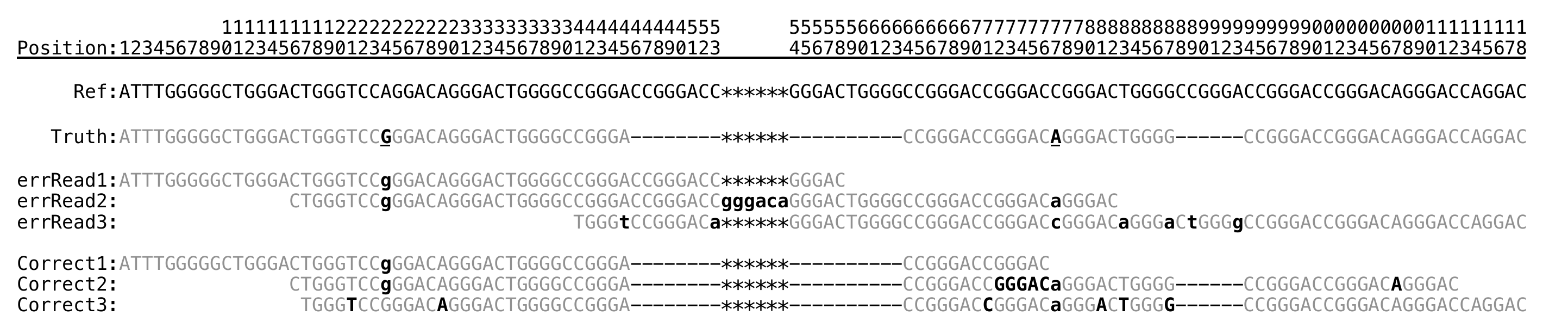}
\caption{Example of misalignment around chr1:26608841 in CHM1. The truth allele
is derived from local assembly. Three erroneous read alignments and their
correct alignments are shown below it. Each of the three reads is an exact
substring of the truth allele, but their alignments are different.  The first
read `errRead1' is aligned without gaps as the 3'-end of the read is a
substring of the 18bp deletion.  Read `errRead2' is aligned with a 6bp
insertion as this alignment is better than having two long deletions. Read
`errRead3' is also aligned without gaps but with seven mismatches. It is
possible for an aligner to find its correct alignment given a small gap
extension penalty. On this example, Bowtie2 did not align any reads with gaps.
BWA-MEM aligned four reads correctly. Except HaplotypeCaller which locally
assembled reads, other callers all called multiple heterozygotes around this
region.}\label{fig:realign}
\end{figure*}

\subsubsection{Realignment errors}
When mapping a read to the reference genome, a read mapper chooses the optimal
pairwise alignment for each read independent of others. For reads mapped to the
same region, the combination of optimal pairwise alignments does not always
yield the optimal multi-alignment of reads. If a variant caller simply trusts
the suboptimal multi-alignment, it may produce false variants or genotypes
(Figure~\ref{fig:realign}). Therefore, more recent variant callers, including
HaplotypeCaller, Platypus and FreeBayes in this study, heavily rely on
realignment for both SNP and INDEL calling.

However, with our manual review, we found that variant callers often failed to
produce the optimal realignment in LCRs. About 50--70\% of the reviewed
$>$1bp heterozygous INDELs from CHM1 can be corrected away with better
realignment. Without the thorough understanding of the very details of the
realignment process, we are unable to explain why the callers fail even on some
obvious cases. Nonetheless, as we can often manually derive a better
multi-alignment, it is possible that a good realignment algorithm may replace
our manual work and achieve higher accuracy than all the tools in our
evaluation.

In the process of manual review, we found local assembly with fermi is
frequently more effective than the INDEL callers, which may be due to the
independence of the reference sequence, the requirement of long-range
consistency and the more powerful topology based error
cleaning~\citep{Zerbino:2008uq}. Some difficult errors such as
Figure~\ref{fig:realign} are trivial to resolve with local assembly.

\begin{figure}
\includegraphics[width=.49\textwidth]{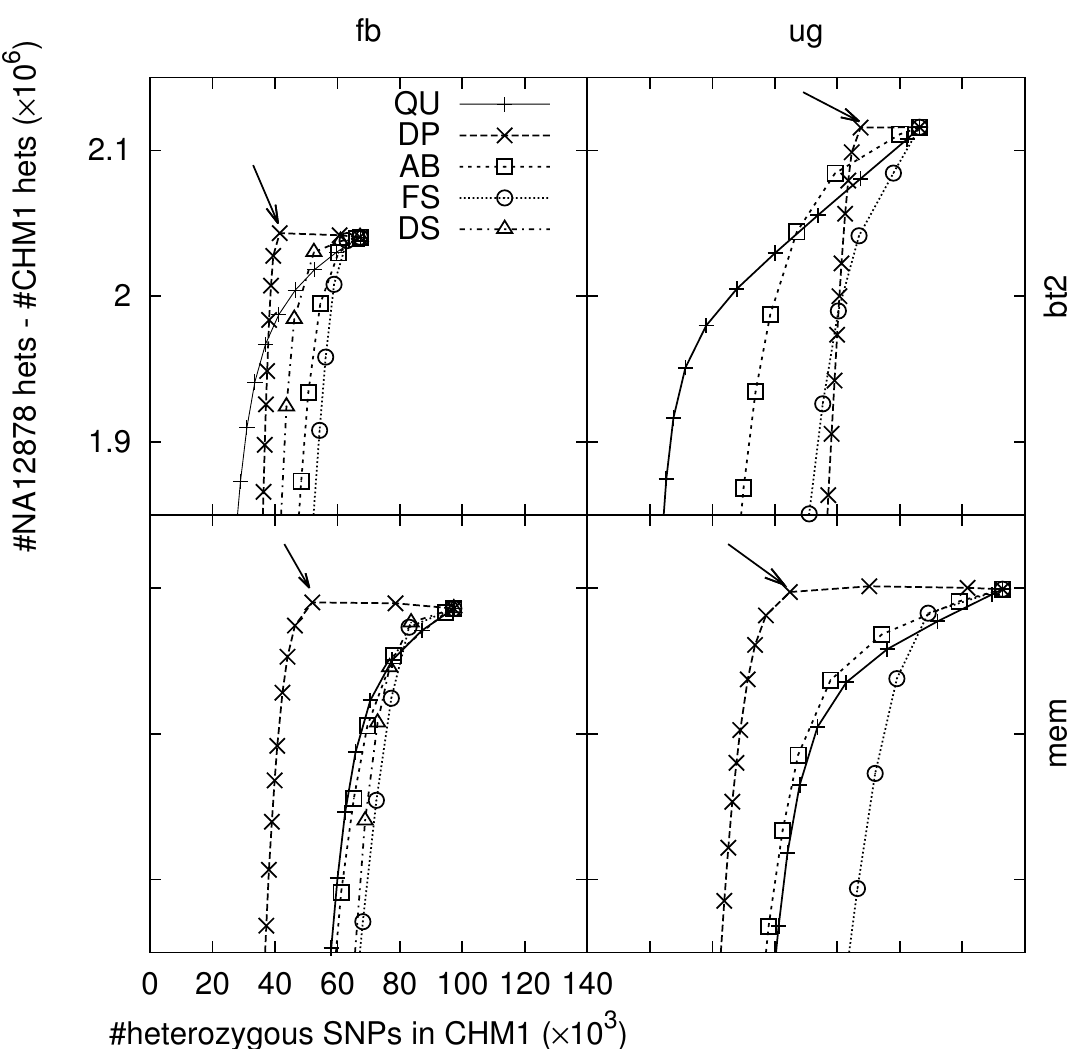}
\caption{Effect of filters after removing variants in low-complexity regions.
Each filter is associated with one value.  For each filter, the number of
heterozygous SNPs called from CHM1 and NA12878 are counted accumulatively from
the most stringent threshold on the filter value to the most relax threshold.
Thresholds are chosen such that they approximately evenly divide variants into
100 bins. Each chosen threshold yields a point in the plot. An arrow points
to a point on the DP curve when the corresponding read depth is right above
$d+4\sqrt{d}$, where $d$ is the mean read depth across called variants.}\label{fig:roc} \end{figure}

\subsection{The maximum read depth filter}

\subsubsection{The effectiveness of the max depth filter}
Other filters require a threshold on a single value. To study
which filter is more effective, we used a ROC-like plot,
Figure~\ref{fig:roc}. In this figure, the X-axis, the number of heterozygous
SNPs in CHM1, is proportional to the false positive rate; the Y-axis, the
difference of the number of heterozygous SNPs between NA12878 and CHM1, serves
as a proxy to the sensitivity. Similar to a standard ROC plot, a curve closer
to the top left corner implies a better classifier of errors.

Figure~\ref{fig:roc} implies that the maximum depth filter is the most
effective against false heterozygotes, especially those found from the BWA-MEM
alignment. On our data with depth $d\approx50$, a maximum depth threshold
between $d+3\sqrt{d}$ and $d+4\sqrt{d}$ removes many false positives with
little effect on the sensitivity. These false positives are mostly caused by
copy number variations (CNVs) or paralogous sequences not present in the human
reference genome.

\subsubsection{The difference between Bowtie2 and BWA-MEM alignment}
It is clear that Bowtie2 is less affected by the presence of CNVs and an
incomplete genome (Figure~\ref{fig:hist} and~\ref{fig:roc}). With manual
review, it seems to us that in comparison to BWA-MEM, Bowtie2 tends to give the
same alignment a lower mapping quality when the read has other suboptimal hits.
At the same time, missing paralogous sequences from the reference genome are
often associated with existing segmental duplications in the reference genome.
Therefore, Bowtie2 is more likely to correctly give a low mapping quality to a
read from these paralogous sequences. As variant callers usually distrust
mismatches on alignments with low mapping quality, their calls from the Bowtie2
alignment are less susceptible to CNVs or an incomplete reference genome.

However, being conservative on the mapping quality estimate may lead to more false
negatives. For example, we found a read pair having one mismatch around
13.7Mbp in chr1 but two mismatches around 13.5Mbp. Both Bowtie2 and BWA-MEM
mapped the ends of the pair at the same positions. Bowtie2 gives the pair a
mapping quality 6, while BWA-MEM gives a mapping quality 27. The similar
scenario happens to the other reads mapped to this region. As a result, a SNP
is called from the BWA-MEM alignment, but not from the Bowtie2 alignment.
Variants callers usually call more variants from the BWA-MEM alignment
(Figure~\ref{fig:hist}), many of which are located in segmental duplications.

Another difference, not relevant to the mapping quality, comes from the alignment
around long INDELs. HaplotypeCaller always called more $\ge$15bp INDELs from
the BWA-MEM alignment (data not shown). Other callers made three times as many $\ge$15bp
deletion calls from the BWA-MEM alignment, either in LCRs or not, and called
40\% more insertions outside LCRs. Interestingly, except HaplotypeCaller,
others called more $\ge$15bp insertions from the Bowtie2 alignment in LCRs
instead. We have not found a good explanation to the apparently conflictive
observations.

\subsubsection{An alternative to the maximum depth filter}
While the maximum depth filter is effective against false heterozygotes, it is
only applicable to high-coverage data with uniform read depth. It
does not work with exome sequencing data, or is not powerful on data with
shallow coverage.

To overcome the limitation, we derived an alternative filter. We obtained
unfiltered SAMtools SNP calls from the 1000 Genomes Project and computed the
inbreeding coefficient and the Hardy-Weinberg P-value using genotype
likelihoods~\citep{Li:2011ab}. We extracted sites satisfying: 1) the reported
read depth above 25,000; 2) the inbreeding coefficient less than zero; 3) the
P-value below $10^{-10}$. We then clustered the sites within 10kb into regions.
These regions are susceptible to common CNVs or artifacts in the reference
genome. We call this filter as the Hardy-Weinberg filter, or HW in brief.

On CHM1, the HW filter is almost as effective as the maximum depth filter.
It could be a valid alternative when the maximum depth filter cannot be
applied.  However, the derivation of the HW filter requires multiple
thresholds and depends on populations, the mapper (BWA) and the caller
(SAMtools). Therefore, we decided to use the much simpler maximum depth filter
here.

\subsection{Other filters}
The remaining filters, including AB, DS, FS and QU (Section~\ref{sec:flt}), can
filter additional false heterozygous called from CHM1, but their effectiveness
varies with call sets. It is also difficult to find the optimal thresholds on
these filters as they affect both the false negative rate and the false
positive rate. In the end, we arbitrarily chose reasonable thresholds based on
the ROC-like curves (Figure~\ref{fig:roc}), which may not be optimal for all
call sets.

\subsection{Effect of PCR duplicates}
20\% of CHM1 data are discarded in our analysis due to PCR duplicates. We have
also tried variant calling with mem:hc without the MarkDuplicates step. Before
filtering, this approach yields 3\% more heterozygous SNPs and 12\% more
heterozygous INDELs, suggesting INDELs are more susceptible to PCR artifacts
than SNPs. After filtering, the total numbers of SNPs and INDELs are about the
same with or without duplicates.

\begin{figure}[!hbtp]
\centering
\includegraphics[width=.35\textwidth]{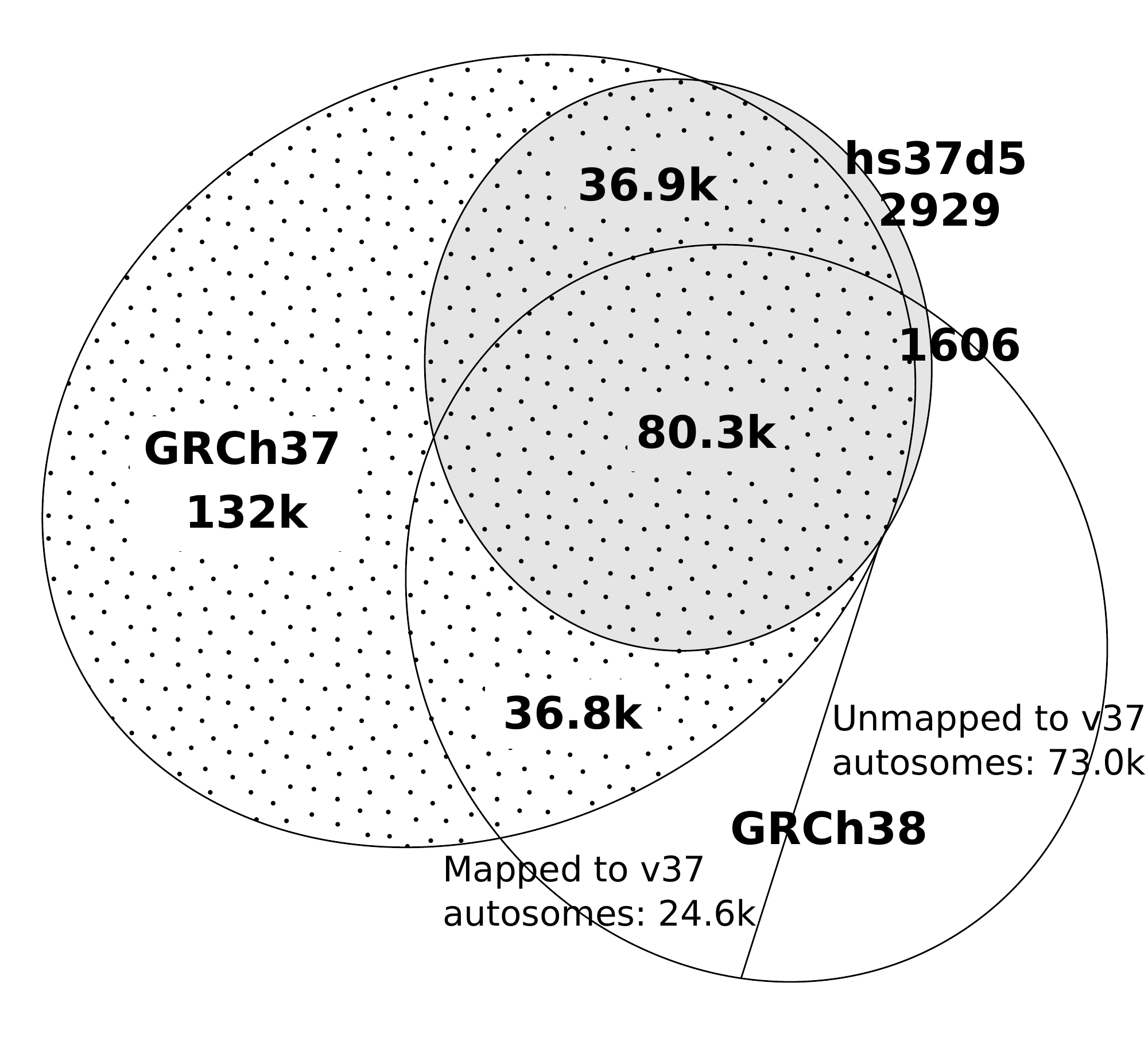}
\caption{Relationship of CHM1 heterozygous SNPs called from mappings to
different reference genomes. CHM1 reads were mapped with BWA-MEM. Autosomal
SNPs were called with GATK HaplotypeCaller and passed the low-complexity
filter. Heterozygous calls from GRCh38 were lifted to GRCh37 with the liftOver
tool from UCSC under the default setting.}\label{fig:ref}
\end{figure}

\subsection{Effect of the reference genome}\label{sec:ref}
In this work, we mapped reads to hs37d5, the reference genome used by the 1000
Genomes Project. This reference genome contains extra 35.4Mb sequences present
in several de novo assemblies but likely to be missing from the primary
assembly of GRCh37. These sequences are supposed to attract many mismapped
reads, so are called as \emph{decoy} sequences.

We have also mapped the CHM1 reads to the GRCh37 and GRCh38 primary assemblies
and called variants. The number of homozygous non-LC SNPs called from each
reference is close: 2.408, 2.405 and 2.412 million from GRCh37, hs37d5 and
GRCh38, respectively. However, the numbers of heterozygous SNPs/INDELs are
distinct (Figure~\ref{fig:ref}).  We called twice as many heterozygotes from
GRCh37 in comparison to hs37d5.  This indicates that the 35.4Mb decoy sequences
indeed attracted many mismapped reads and consequently improves the variant
calls in chromosomal regions.  GRCh38 further resolves 39.8k (=36.9k+2909)
heterozygotes called from hs37d5.  However, it also retains 36.8k heterozygotes
called from GRCh37 but not from hs37d5. Intriguingly, GRCh38 further adds 24.6k
autosomal heterozygotes not called from GRCh37 or hs37d5. We are unclear of the
source of these false heterozygous SNPs. In general, we conclude that hs37d5
and GRCh38 are more complete than GRCh37.

\section{Discussions and Conclusions}
A distinct feature of our works is the use of a haploid human sample, CHM1,
from which heterozygous calls are supposed to be errors. This allows us to unbiasedly
investigate the causal artifacts and to experiment effective filters with the
diploid NA12878 data set as a positive control.

When we called SNPs and INDELs from CHM1, we were surprised to find 10\% of raw
variant calls were heterozygotes. Honestly, our immediate reaction was that
CHM1 was not truly haploid. However, after careful analysis, we have convinced
ourselves that the heterozygosity of CHM1 should be of an order of magnitude
lower than the raw error rate of variant calling. The vast majority of
heterozygotes are calling errors. In the raw call set, we usually see an error
per 10--15kb.

It was also to our surprise that the low-complexity filter is the most
effective against false heterozygotes, especially short INDELs. Although we
knew that INDEL errors may be introduced by PCR during sample preparation, we
underestimated its substantial effect. We were also unware that realignment of
INDELs in LCRs remains a great challenge even after the many existing
efforts in this
direction~\citep{Homer:2010aa,Li:2011kx,Albers:2011aa,Narzisi:2014aa}. Without
the suggestion from Peter Sudmant (personal communication), we would not have
tried this
filter.

Before we understand and resolve the issues in variant calling in LCRs,
it might be better to filter out all variants overlapping these regions.
Although over 50\% of single-sample INDEL calls fall in LCRs
(Figure~\ref{fig:hist} and~\ref{fig:venn-NA12878}), only 1.25\% of autosomal
INDELs in the ClinVar database (http://clinvar.com) overlap with LCRs -- most
INDELs in LCRs have unknown clinical functionality. For certain applications, it might be safe to drop or downweigh
these difficult calls.

Outside LCRs, different call sets usually agree well with each other
if the same set of filters are applied (Figure~\ref{fig:venn-NA12878}). Based
on Figure~\ref{fig:hist}, we estimate that a caller usually makes a wrong call
per 100--200kb without significant compromise on the sensitivity, similar to
the previous estimates~\citep{Bentley:2008cr,Nickles:2012aa}. Many of these
errors are likely to be systematic. In the context of somatic or \emph{de novo}
mutation discovery by sample contrast, systematic errors will appear in all
samples.  They will not lead to false mutation calls, fortunately.

A simple method to improve the variant accuracy is to use two distinct pipelines,
take the intersection of the raw calls and then apply caller-oblivious filters
to derive the final call set. As callers agree well on post-filtered sites
(Figure~\ref{fig:venn-NA12878}) but badly on false positives
(Figure~\ref{fig:venn-CHM1}), we should be able to remove most errors without
much hit to the sensitivity. Such a consensus approach has been applied to
cancer data with limited success~\citep{Lower:2012aa,Goode:2013aa}. Without
subclonal mutations, it should be much more effective on the variant discovery
from normal samples.

Finally, the advances in sequencing technologies lead the development of
algorithms. We are heavily relying on mapping based variant calling because
with very short reads or at low coverage, the traditional
assembly-and-mapping approach would not work. With increased read
lengths and decreased sequencing cost, we might go back to \emph{de novo}
assembly. An assembly does not only encode small variants, but also
retains large-scale structural variations and is free of the artifacts in the
reference genome. Another possible direction which we mentioned four years
ago~\citep{Li:2010kx} is to map sequence reads to the ensemble of 
multiple genomes. Recently, there have been significant progress towards this
goal~\citep{DBLP:journals/corr/abs-1010-2656,Paten:2014aa}, but a practical solution is yet to
be concluded.

\section*{Acknowledgement}
We are really grateful to Richard Wilson and his team who sequenced the CHM1 cell line
and granted us the permission to use the data in this study. We also thank the
1000 Genomes Project analysis group for the helpful comments and discussions,
and thank Mike Lin and the anonymous reviewers whose comments have helped us to
improve the manuscript.

\paragraph{Funding\textcolon} NHGRI U54HG003037; NIH GM100233

\bibliography{varcmp}
\end{document}